\documentclass[twocolumn]{aastex63}

\usepackage{float}
\graphicspath{{./}{figures/}}

\received{January 13, 2021}
\revised{February 1, 2021}
\accepted{February 1, 2021}
\submitjournal{ApJ lett.}

\shorttitle{A self-consistent  simulation of a SIR event}
\shortauthors{Wijsen et al.}

\newcommand{\vect}[1]{\mathbf{#1}}
\renewcommand{\vec}{\vect}

\begin{document}

\title{A self-consistent simulation of proton acceleration and transport near a high-speed solar wind stream}

\correspondingauthor{Nicolas Wijsen}
\email{nicolas.wijsen@kuleuven.be}

\author[0000-0001-6344-6956]{Nicolas Wijsen}
\affiliation{Centre for mathematical Plasma-Astrophysics, Department of Mathematics, KU Leuven, Celestijnenlaan 200B, B-3001 Leuven, Belgium}

\author[0000-0002-7676-9364]{Evangelia Samara}
\affiliation{Royal Observatory of Belgium, Brussels, Belgium}
\affiliation{Centre for mathematical Plasma-Astrophysics, Department of Mathematics, KU Leuven, Celestijnenlaan 200B, B-3001 Leuven, Belgium}

\author[0000-0003-1539-7832]{\`Angels Aran}
\affiliation{Department of Quantum Physics and Astrophysics, Institute of Cosmos Sciences (ICCUB), Universitat de Barcelona (UB-IEEC), Spain}

\author[0000-0002-3176-8704]{David Lario}
\affiliation{NASA, Goddard Space Flight Center, Heliophysics Science Division, USA}

\author[0000-0003-1175-7124]{Jens Pomoell}
\affiliation{Department of Physics, University of Helsinki, Helsinki, Finland}

\author[0000-0002-1743-0651]{Stefaan Poedts}
\affiliation{Centre for mathematical Plasma-Astrophysics, Department of Mathematics, KU Leuven, Celestijnenlaan 200B, B-3001 Leuven, Belgium}
\affiliation{Institute of Physics, University of Maria Curie-Sk{\l}odowska, Lublin, Poland}

\begin{abstract}
%The acceleration of energetic particles observed in association with the passage of solar wind stream interaction regions (SIRs) may occur locally near the observer or at distant ($>$1 au) helioradii where SIR shocks form. 
 
Solar wind stream interaction regions (SIRs)  are often characterised by energetic ion enhancements. 
The mechanisms accelerating these particles, as well as the locations where the acceleration occurs, remain debated.
Here, we report the findings of a simulation of a SIR event observed by Parker Solar Probe at $\sim$0.56~au and the Solar Terrestrial Relations Observatory-Ahead at $\sim$0.95~au in September 2019 when both spacecraft were approximately radially aligned with the Sun.
The simulation reproduces the solar wind configuration and the energetic particle enhancements observed by both spacecraft.
Our results show that the energetic particles are produced at the compression waves associated with the SIR and that the suprathermal tail of the solar wind is a good candidate to provide the seed population for particle acceleration. 
The simulation confirms that the acceleration process does not require shock waves and can already commence within Earth's orbit,  with an energy dependence on the precise location where particles are accelerated. 
The three-dimensional configuration  of the solar wind streams strongly modulates the energetic particle distributions, illustrating the necessity of advanced models to understand these particle events.
\end{abstract}

\section{Introduction}
The interaction of a high-speed solar wind stream (HSS) with slower solar wind ahead results in the formation of a stream interaction region \citep[SIR;][]{gosling99}.
SIRs observed at 1~au are usually bounded by forward and reverse compression waves (FCW and RCW). 
The FCW accelerates and compresses the slow solar wind propagating in front of the HSS, whereas the  RCW decelerates and compresses the fast solar wind of the HSS.
 At larger heliocentric distances, such pressure waves commonly steepen into forward and reverse shocks (FS and RS) \citep{gosling99}.

It was originally suggested that recurrent energetic ion intensity enhancements seen at 1~au in association with HSSs were produced by these outer forward--reverse shock pairs \citep[e.g.,][]{barnes76,fisk80}.
Intensity radial gradients \citep{vanhollebeke78} and flow directions \citep{marshall78} of the SIR-associated energetic ions were consistent with the origin of the particles at several~au \citep{mason99}.
However, \cite{giacalone02} realised  that compression waves can already induce a first-order Fermi acceleration process, without needing to steepen into shock waves. 
This happens when the spatial extent of the compression  is significantly smaller than the mean free path of the particles across the wave, such that the particles experience the compression wave much like a shock \citep{giacalone02}. In addition, a second-order Fermi acceleration process may also contribute to the local ion acceleration in SIRs \citep[e.g.,][]{richardson85,schwadron96,schwadron20b}.

An important question remains with regard to the dominant source  of the SIR-associated energetic ions detected at 1~au, namely whether they originate  from distant shock waves or from close-by compression waves.
Observations from Helios \citep{Porsche75}, Parker Solar Probe \citep[PSP;][]{fox16}, and Solar Orbiter \citep[SolO;][]{muller13}  have shown that SIR energetic ions are detected well within the Earth's orbit \citep[e.g.,][]{vanhollebeke78,mcComas19, desai20,cohen20,allen20, allen20d, allen20c, joyce20,schwadron20b}.
Hence, the study and modelling of SIR particle events at different heliocentric distances is essential in order to address the acceleration and transport of energetic particles. 

In this work, we study a SIR event observed \citep{allen20b} in September 2019 by both PSP (located at ${\sim}0.56$~au) and the Solar Terrestrial Relations Observatory-Ahead (STEREO-A) spacecraft (located at ${\sim}0.95$~au). 
 The close radial alignment between STEREO-A and PSP at that time makes this event especially interesting because the evolution of the coronal hole responsible for the HSS observed by both spacecraft is minimal \citep{allen20b}.    
To understand this SIR event in depth, a realistic scale-bridging model of the solar wind and  energetic particle acceleration and transport is necessary. 
In this work, this is achieved by using the combination of two novel models, namely, the EUropean Heliospheric FORcasting Information Asset  \citep[EUHFORIA;][]{pomoell18} and the PArticle Radiation Asset Directed at Interplanetary Space Exploration \citep[PARADISE;][]{wijsen19a,wijsenPHD20}. 
EUHFORIA is a data-driven magnetohydrodynamic (MHD) model of the solar wind, whereas PARADISE simulates energetic particle distributions in the solar wind provided by EUHFORIA, using a quasi-linear \citep{jokipii66} approach to capture the interaction between solar wind turbulence and energetic particles. 
We illustrate how EUHFORIA+PARADISE  captures, in a self-consistent manner: (1)~the complex HSS structure, (2)~the formation of the SIR, (3)~the acceleration of protons at both the FCW and the RCW, and (4)~the propagation of the energetic protons into the inner heliosphere. 
 We show how the spatial distribution of SIR particles varies as the compression waves tend to steepen at large helioradii. Whereas low-energy particles
($\lesssim 0.2$ MeV) can be produced at small helioradii, the acceleration  of more energetic particles requires steeper compression waves.

\section{Results}

\subsection{The Solar Wind}
\begin{figure*}
    \centering
    \includegraphics[width=0.90\textwidth]{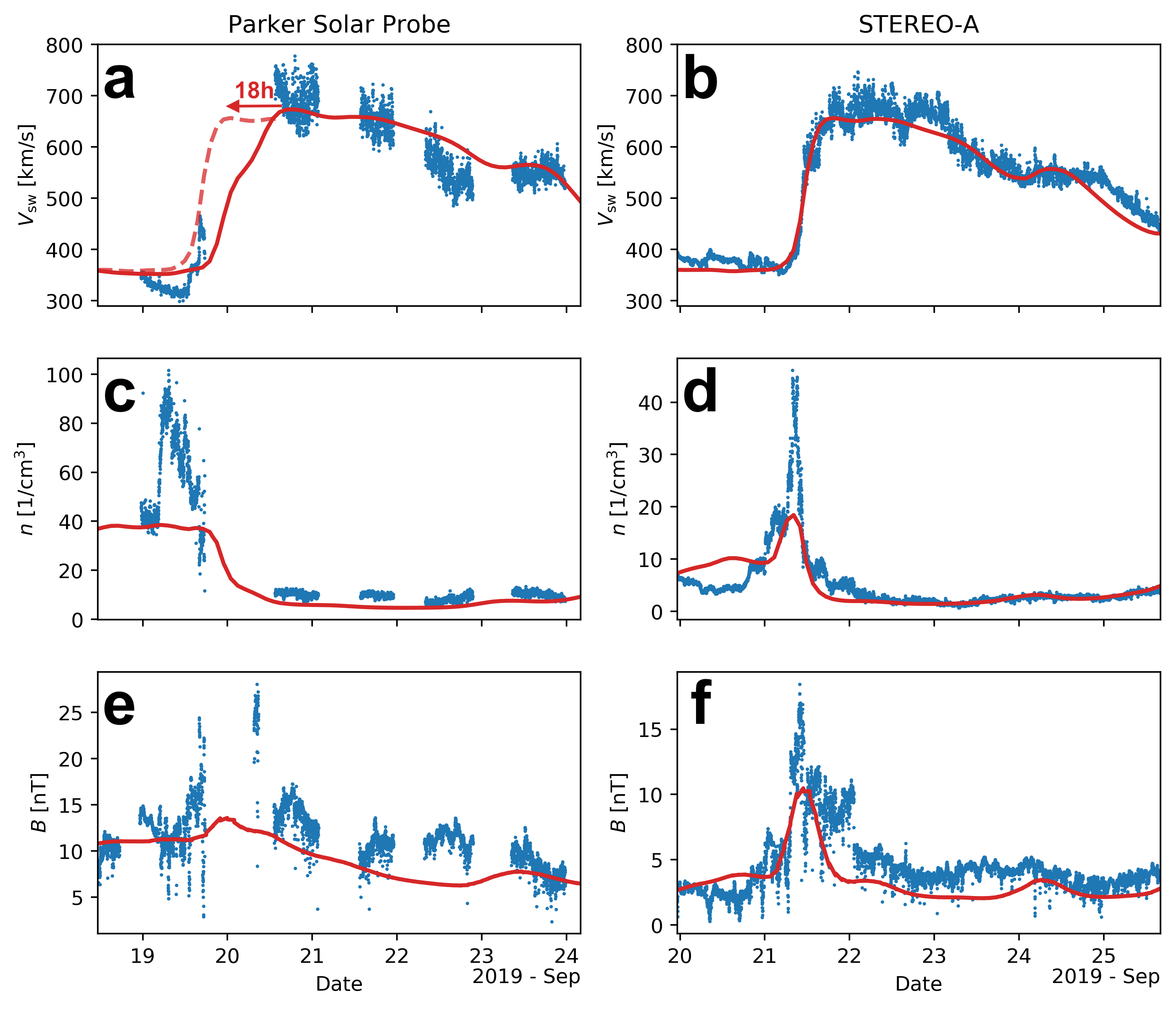}
    \caption{Comparison between the proton solar wind speed (a--b), density (c--d), and magnetic field magnitude (e--f) observed (blue dots) by PSP (left) and STEREO-A (right) with the EUHFORIA simulation results (red solid lines).  
    The dashed line in panel~(a) shows the simulated HSS onset observed at STEREO-A shifted earlier in time by 1.77 days \citep{allen20b} to account for corotation.}
    \label{fig:sw_profiles}
\end{figure*}

\begin{figure*}
\begin{tabular}{c c}
\includegraphics[width=0.45\textwidth]{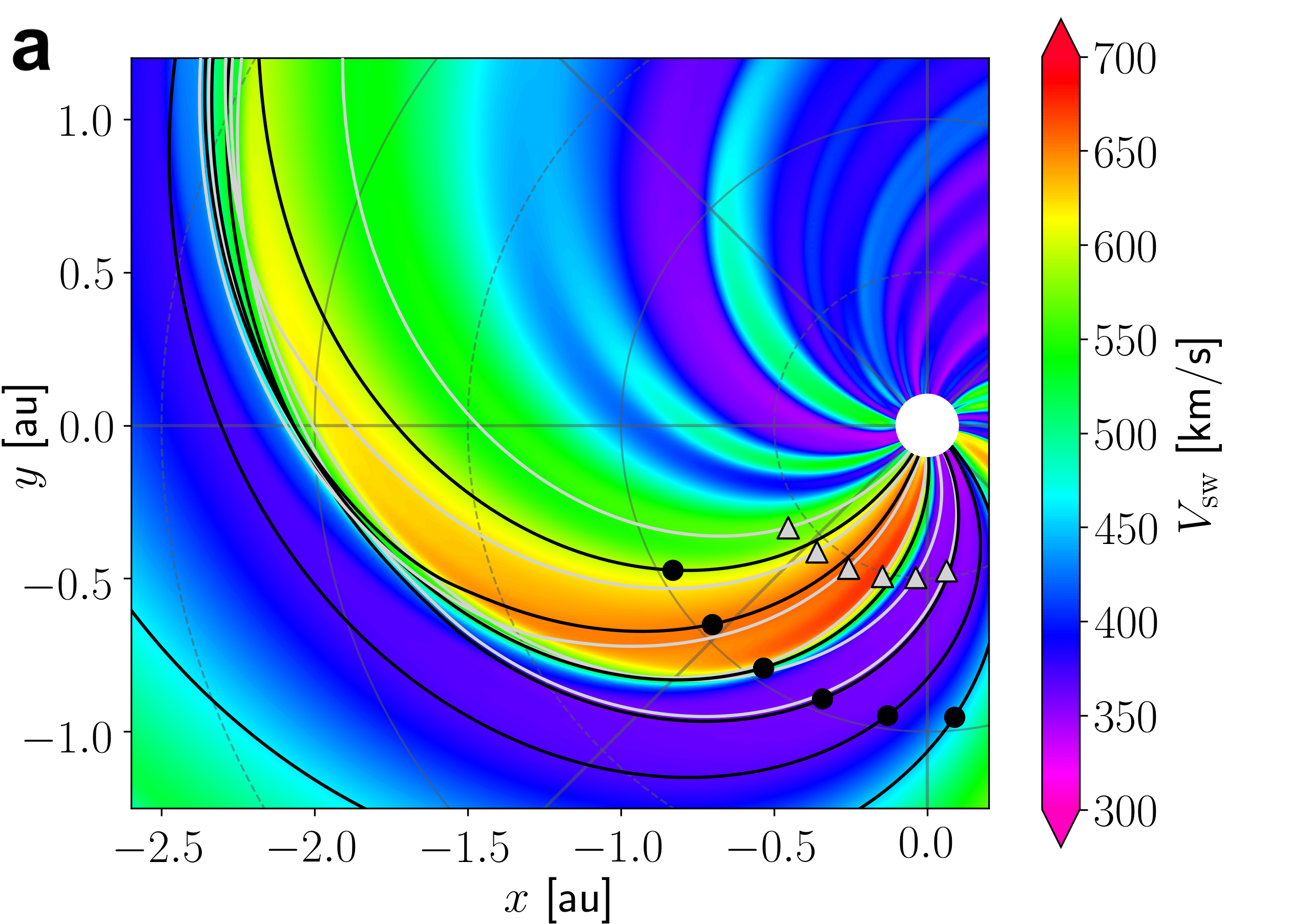} 
&
\includegraphics[width=0.45\textwidth]{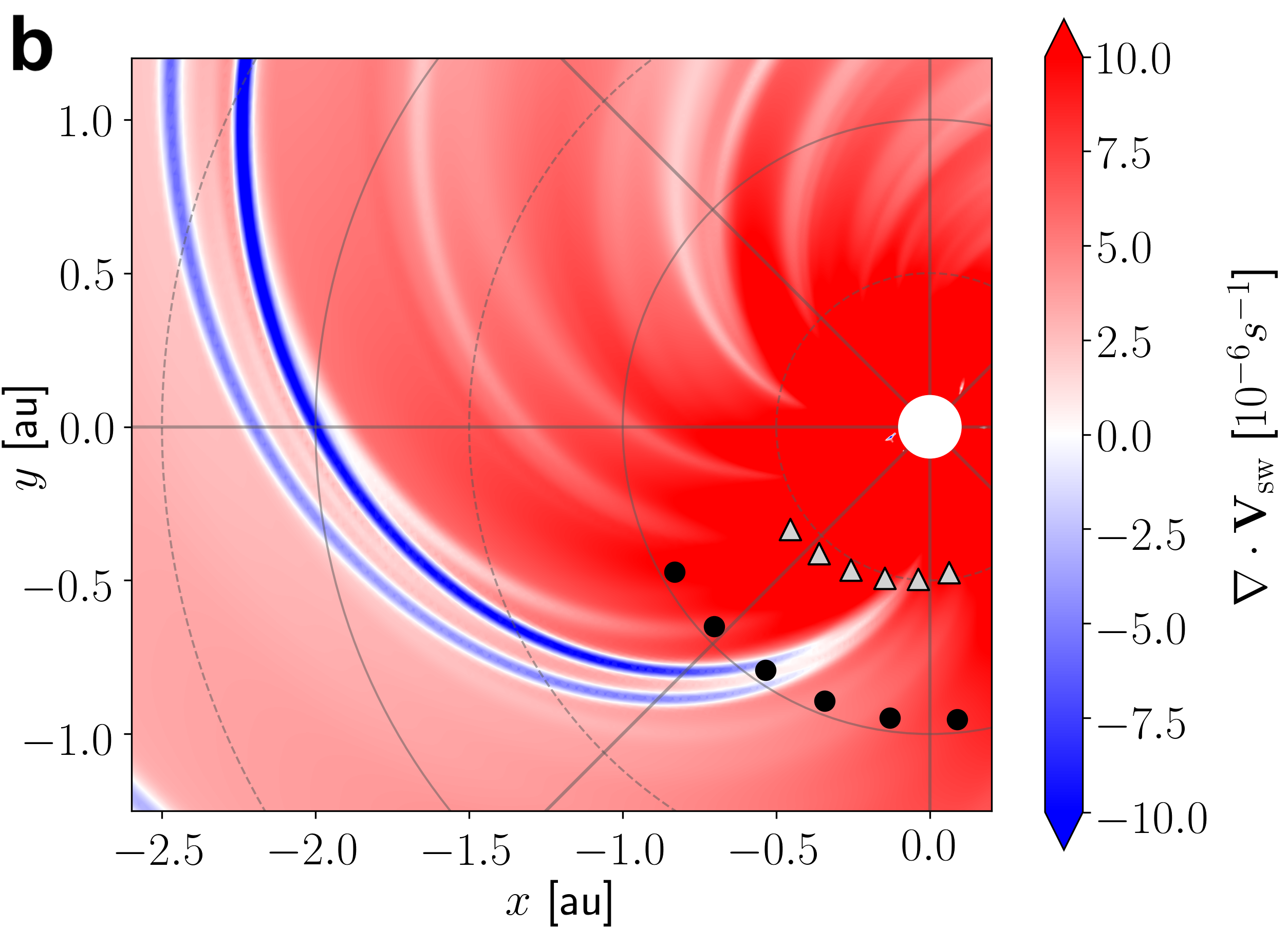}
\\
\includegraphics[width=0.45\textwidth]{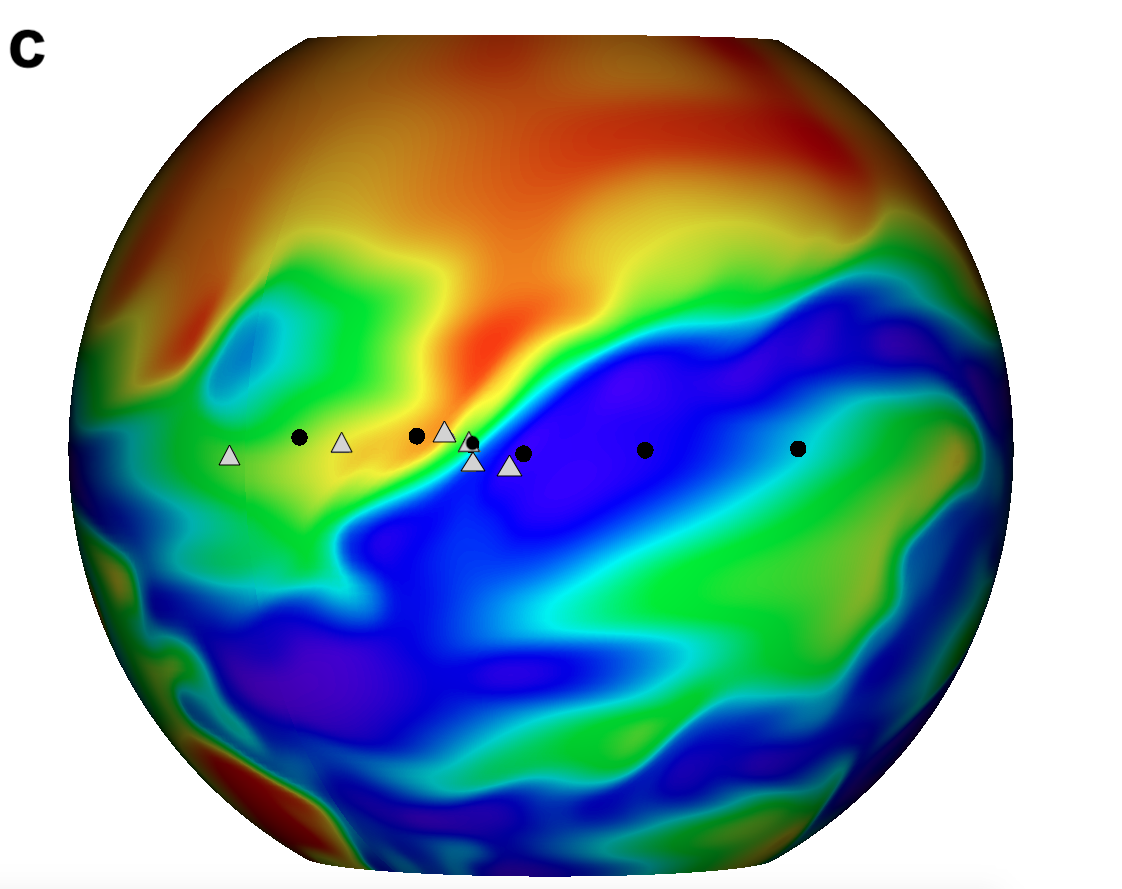}
& \includegraphics[width=0.45\textwidth]{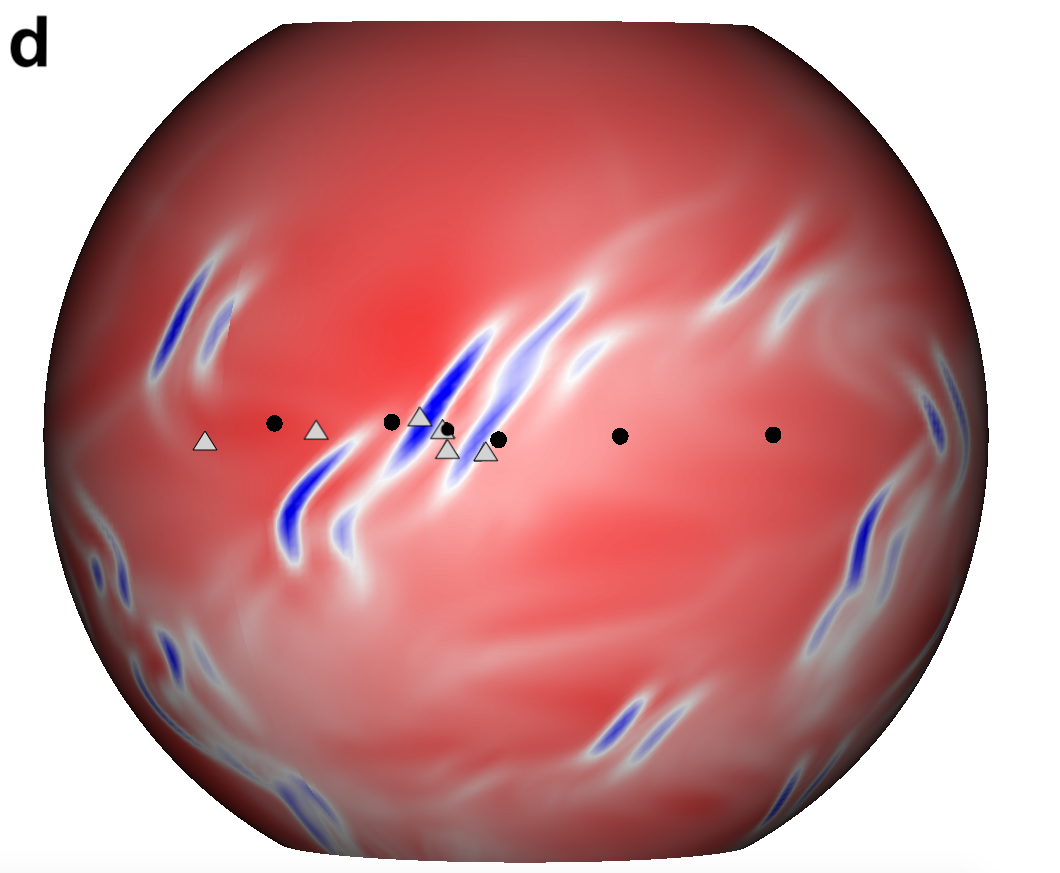}
\end{tabular}
\caption{Simulated solar wind speed  (panels a and c) and the divergence of the solar wind velocity  (panels b and d) at HEEQ latitude  $\vartheta = 2.3^\circ$ (upper row) and at the radial distance $r = 1.5$~au (lower row). The  symbols in panels~a and~b  represent the location of STEREO-A (black dots) and PSP (grey triangles) at a 24h cadence, ranging from 18 Sep 12:00~UT to 23 Sep 12:00~UT (from right to left). The dots and triangles in~c and~d represent the corresponding magnetic connection of STEREO-A and PSP, respectively
}\label{fig:sw_colormaps}
\end{figure*}

Figure~\ref{fig:sw_profiles} shows the solar wind bulk speed, proton number density, and magnetic field magnitude observed by PSP (left) and STEREO-A (right). 
The gaps in the PSP data correspond to time periods when the instruments on board the spacecraft were powered off. 
The results of the data-driven EUHFORIA solar wind simulation are shown in red. The precise As can be seen,  the EUHFORIA simulation succeeds at reproducing the observed speed profile at STEREO-A, with the simulation capturing the amplitude of the HSS as well as
the decreasing speed profile in the rarefaction region behind the HSS. 
Reproducing the speed profile of the solar wind is essential when studying energetic particle populations.
This is because the frozen-in condition of the solar wind magnetic field implies that the large-scale magnetic field connectivity is largely dominated by the solar wind plasma flow. 

 The arrival of the HSS at PSP occurred in the middle of a data gap (Fig.~\ref{fig:sw_profiles}a). 
The EUHFORIA model (red solid line) predicts a solar wind speed increase more gradual than what the available PSP data suggests. 
This simulated gradual increase translates into a small magnetic field increase at the SIR in comparison with the observations (cf. Fig.~\ref{fig:sw_profiles}e).
STEREO-A data suggests the presence of a developing RS on 22 Sep 01:35~UT 
 \citep{allen20b} as shown by the abrupt decrease in magnetic field magnitude.
%The existence of this discontinuity at 0.95~au indicates that the solar wind speed gradient at PSP should already be significant.
\cite{allen20b} found that PSP solar wind data aligns well with STEREO-A data when the latter is shifted 1.77 days in time to account for corotation.
%This suggests that most likely the actual speed increase at PSP was steeper than the simulation result.
The dashed line in Fig.~\ref{fig:sw_profiles}a  shows the simulated speed profile observed by STEREO-A, shifted 1.77 days earlier in time.
For such a more abrupt speed increase, the maximum speed at PSP would be obtained 18h earlier than in the current simulation.
The gradual speed increase at PSP in the simulation results partly from an overestimation of the solar wind speed just ahead of the HSS (see Fig.~\ref{fig:sw_profiles}a).
The existence of a strong speed gradient in the solar wind at 0.5~au suggests a very abrupt transition between the slow and fast solar wind source regions.

Figure~\ref{fig:sw_colormaps} displays the three-dimensional structure of the HSS.
 Figure~\ref{fig:sw_colormaps}a shows the solar wind speed $V_{\rm sw}$ at the constant latitude $\vartheta = 2.3^\circ$ (using the Heliocentric Earth Equatorial (HEEQ) coordinate system), which is approximately  the latitude of PSP and STEREO-A during the SIR passage. 
The symbols indicate the location of STEREO-A (black dots) and PSP (grey triangles) at different times. The black lines are interplanetary magnetic field (IMF) lines crossing the spacecraft and projected on the $\vartheta = 2.3^\circ$ plane.
The apparent crossing of these IMF lines inside the SIR is solely a projection effect.
The non-zero latitudinal component of the IMF arises mainly due to the deflection of the magnetic field at the FCW and the RCW \citep{wijsen19b}. 

Figure~\ref{fig:sw_colormaps}c shows $V_{\rm sw}$ at a constant helioradius $r=1.5$~au. 
The HSS appears as a southward extension of the fast solar wind originating from the northern polar coronal hole. 
The magnetic connection of PSP and STEREO-A at 1.5~au is indicated by the grey triangles and the black dots and at the same times as in Fig.~\ref{fig:sw_colormaps}a.
Both spacecraft connect to similar regions of the SIR, yet for a different time duration. 
PSP has a magnetic connection within the SIR at 1.5~au for approximately three full days (19 Sep 12:00 to 21 Sep 12:00), whereas STEREO-A is only connected for ${\sim} 1.5$ days (20 Sep 12:00 to 22 Sep 00:00) to the same region. 

Figures~\ref{fig:sw_colormaps}b and~\ref{fig:sw_colormaps}d  show  the divergence of the solar wind velocity $\nabla\cdot\vec{V}_{\rm sw}$, which is a measure of the local plasma compression.  
Most of the solar wind is in an expanding state, characterised by $\nabla\cdot\vec{V}_{\rm sw} > 0$.
These are the regions where particles undergo adiabatic cooling while propagating \citep{ruffolo95}. 
In contrast, the FCW and the RCW bounding the SIR are regions where $\nabla\cdot\vec{V}_{\rm sw} < 0$.   
Both waves are clearly visible in  Fig.~\ref{fig:sw_colormaps}b as two  spiral-shaped blue  regions, with larger negative values of $\nabla\cdot\vec{V}_{\rm sw}$ at the RCW.
Figure~\ref{fig:sw_colormaps}d illustrates the complex latitudinal structure of the SIR, showing that the solar wind contains several additional compressed plasma regions that are all characterised by a clear two-component structure, that is, a FCW followed by a (stronger) RCW. 
Each of these compression regions forms a potential source of accelerated particles. 
However, most of these regions are located well above or below the ecliptic. 
Figure~\ref{fig:sw_colormaps}d shows that, after passing the RCW, the magnetic connection of both STEREO-A and PSP skim the northern edge of another RCW, associated with the southern edge of the HSS. 
As discussed below, this second RCW may be responsible for the small particle intensity enhancements seen at STEREO-A on 23 Sep (see Fig.~\ref{fig:STA-intensities}).  

\subsection{Energetic Particles at STEREO-A}
\begin{figure*}
\begin{tabular}{c c}
\includegraphics[width=0.45\textwidth]{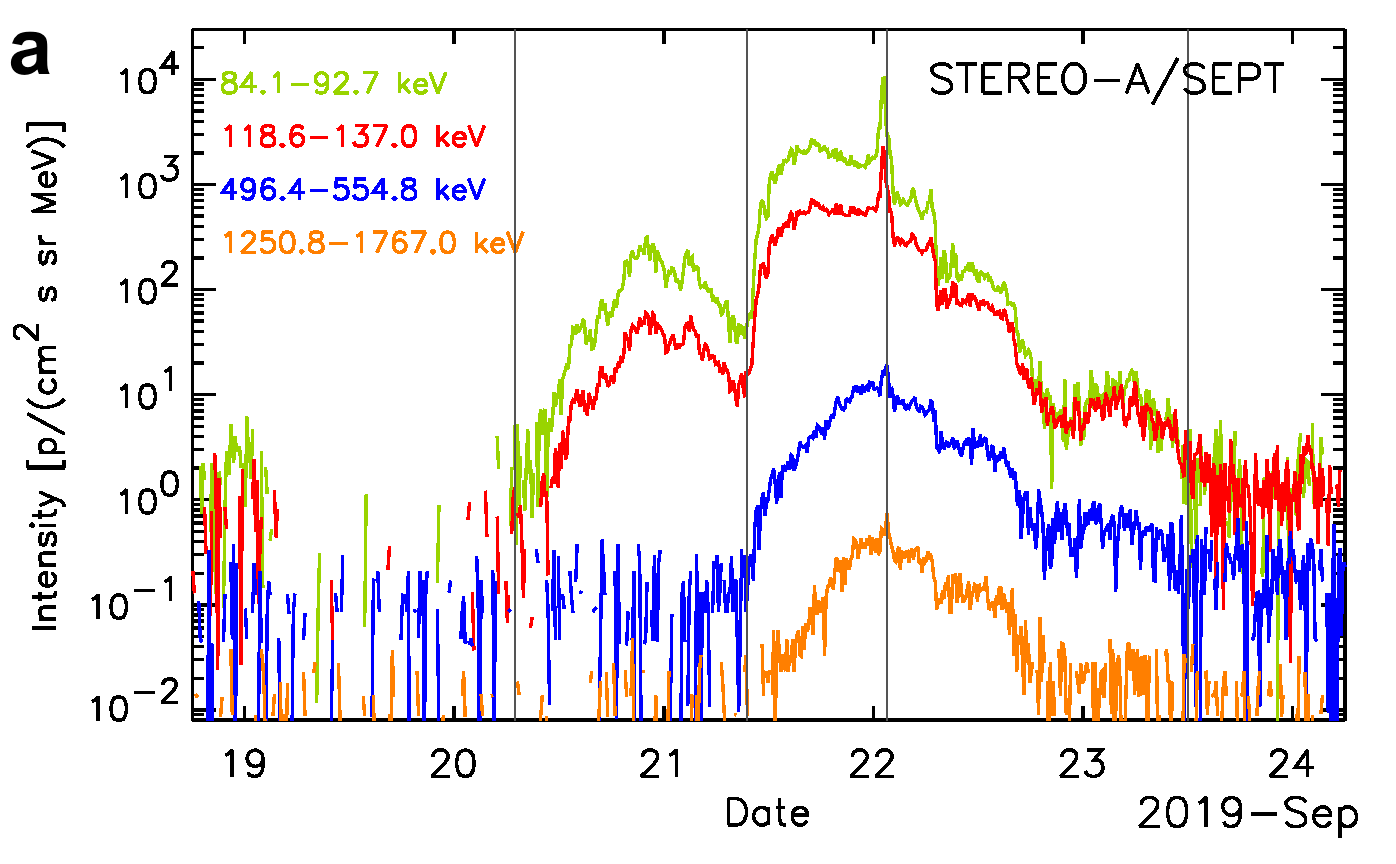} 
&
\includegraphics[width=0.45\textwidth]{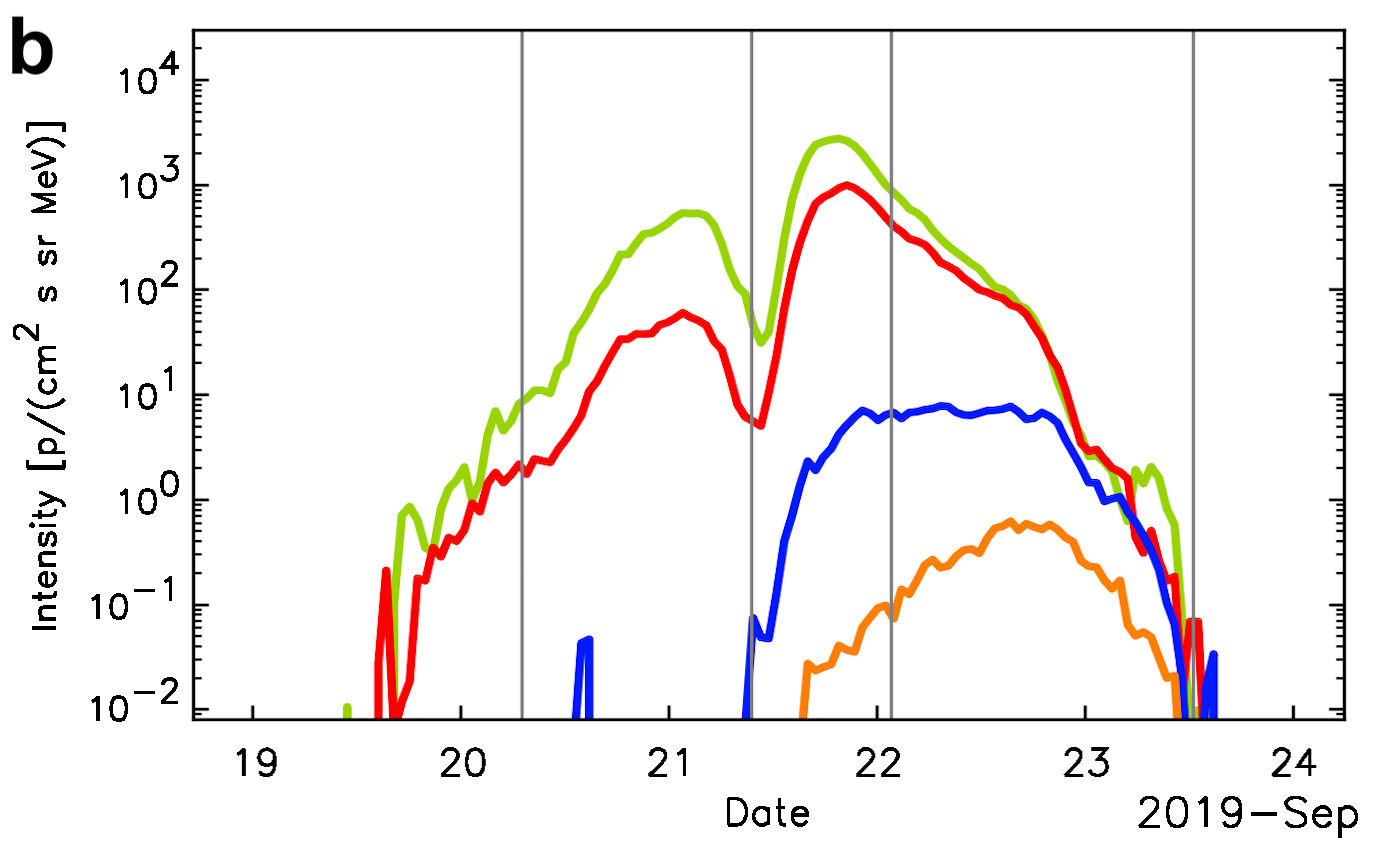}
\end{tabular}
\caption{Observed (left) and simulated (right) omnidirectional ion intensities at STEREO-A. The vertical lines indicate the onset time of the SIR event (Sep 20 09:00~UT), the SI (Sep 21 09:30~UT), the developing RS (22 Sep 01:35~UT), and the stop time of the SIR event (23 Sep 12:00~UT). }\label{fig:STA-intensities}
\end{figure*}
Figure~\ref{fig:STA-intensities}a shows the 10 minute averaged ion intensities observed by the Solar Electron and Proton Telescope  \citep[SEPT;][]{muller-mellin08} on board STEREO-A in four different energy channels, after subtracting the corresponding pre-event intensity values (averaged from Sep 18 00:00~UT to Sep 19 12:00~UT).  
Intensity enhancements at energies below $\sim$200 keV were observed already on Sep 20 09:00~UT, whereas $\gtrsim$500 keV proton intensities did not increase until mid 21 Sep. Low-energy ($\lesssim$500 keV) proton intensities continued to be elevated until 23 Sep 12:00~UT.

To simulate these particle populations, a seed population of 40~keV protons is injected into both the RCW and the FCW bounding the SIR. 
The PARADISE model propagates these protons through the EUHFORIA solar wind, taking into account the effect of small-scale solar wind turbulence on the energetic particles. 
The latter is done using results from quasi-linear theory, which translates the scattering processes of particles by magnetic fluctuations into an anisotropic pitch-angle diffusion process in the reference frame co-moving with the solar wind \citep{jokipii66}.  
This diffusion process drives the particle acceleration in the solar wind regions where $\nabla\cdot\vec{V}_{ \rm sw} < 0$, as it models the interaction of particles with converging scattering centres \citep[e.g.,][]{leRoux12,zank14}.
In addition to pitch-angle diffusion,  a weak spatial diffusion process perpendicular to the IMF is considered. 
The precise set-up of the PARADISE simulation is described in Appendix~\ref{app:paradise}.

Figure~\ref{fig:STA-intensities}b shows the simulated intensity profiles, under the assumption that the protons propagate with constant mean free paths along and across the IMF of $\lambda_\parallel = 0.3$~au and $\lambda_\perp = 10^{-4}$~au, respectively. 
The assumption that $\lambda_\perp/\lambda_\parallel \sim 10^{-3} $ implies that the energetic particles are predominantly propagating along the IMF lines.
Figure~\ref{fig:STA-intensities}b shows that the
PARADISE simulation successfully reproduces several features seen in Figure~\ref{fig:STA-intensities}a, specifically: (1) the onset and ending  times of the SIR event; (2) the double-peaked structure of the intensity-time profiles, with the second peak showing the highest intensities; (3) the soft energy spectrum of the first peak compared to the second peak, with the first peak showing very few protons above ${\sim} 500$~keV; (4) the sudden strong increase in the particle intensities of the two lowest energy channels around 21 Sep 10:00~UT before reaching a second peak; (5) the gradual increase in the two high-energy channels,  hence producing an overall energy spectrum that hardens with the passing of the SIR; (6) the fact that intensity increases are being observed at MeV energies; and (7) the energy spectrum obtained from integrating the intensity over the entire event that has a power-law dependence of $E^{-2.94\pm 0.01}$ in the data and $E^{-2.90\pm 0.02}$ in the simulation. Here, $E$ denotes the particle energy and the power-law fit was performed for the energy range 84.1 -- 1985.3 keV.

The double-peaked profile is a result of the particles being accelerated at either the FCW or the RCW. 
The FCW is weaker than the RCW, and hence less efficient in accelerating particles to high energies.
 Because of the $\nabla\cdot\vec{V}_{\rm sw}$ configuration (Figures~\ref{fig:sw_colormaps}b), 
particle acceleration occurs from ${\sim}$0.8~au onward, 
with the most efficient particle acceleration occurring around ${\sim}$1.5~au. 
As a result, the majority of the accelerated protons detected at STEREO-A in the simulation are observed to be streaming in the sunward direction, even though some low-energy protons ($\lesssim$ 90~keV) are already accelerated within 1~au. 
In addition, since our solar wind simulation underestimates the speed gradient at the location of PSP, it is very likely that, in reality, more particles were already accelerated within 0.8~au.

The observed particle intensities show a sudden spike on 22 Sep 01:35~UT, which coincides with the passage of the developing RS. 
However, in the EUHFORIA simulation, no shock wave is present as the RCW  is still steepening beyond 0.95~au. 
Only at radial distances beyond 1~au does the RCW steepens enough to efficiently accelerate particles to MeV energies. 
This explains why, in the simulation, the MeV energy channel peaks too late compared to the data. 
In contrast, the peak intensities of the two lowest energy channels in  Fig.~\ref{fig:STA-intensities}b coincides with the passage of the simulated FSW and the RSW, indicating that these particles are accelerated locally.

In the simulation, the widths of the compression regions remain larger than the widths of the observed transitions due to the size of the EUHFORIA computational grid. 
However, the  diffusive length scale of the particles across the compression waves is, in the simulation, substantially larger than the width of the compression waves itself. 
As a result, particles accelerate similarly to a diffusive shock acceleration process \citep{giacalone02,wijsen19a}. 
The good agreement between the simulated and observed energy spectra, confirms that shock waves are not required for producing energetic particle enhancements near SIRs.

The intensity dip observed around Sep 21 09:30~UT coincides with the stream interface (SI) of the SIR, that is, the region inside the SIR where the compressed fast solar wind meets the compressed slow solar wind \citep{gosling78}. 
The SI separates the two energetic particle populations that are accelerated at the FCW and the RCW.
The IMF lines directly adjacent to the SI are located inside the developing SIR from small radial distances onward, and as a result, they never cross the FCW or the RCW where particles may accelerate.
In a simulation with zero cross-field diffusion and no particle drifts, the suprathermal particle intensity would thus drop to zero at the SI. 
In contrast, 
if the cross-field diffusion would be more efficient, the two intensity peaks would merge into a single enveloping peak \citep{richardson85,wijsen19c}. 
Since this is not the case in the observed intensity profiles, strong cross-field diffusion near the SI is  inconsistent with the observations.

The observed intensities show a sudden discontinuity on 22 Sep 7:00~UT, which coincides with a rotation in the magnetic field vector (not shown here). This small-scale structure is not resolved by EUHFORIA and hence the same is true for PARADISE.

Finally, Fig.~\ref{fig:STA-intensities}a shows that the particle intensities in the two lowest-energy channels had an additional  small third increase on 23 Sep~01:00 UT. 
In the simulation, this date corresponds to the time when the magnetic connection of STEREO-A is skimming the northern edge of the second RCW shown in Fig.~\ref{fig:sw_colormaps}.

\subsection{Energetic Particles at PSP}
\begin{figure*}
    \centering
    \includegraphics[width=0.6\textwidth]{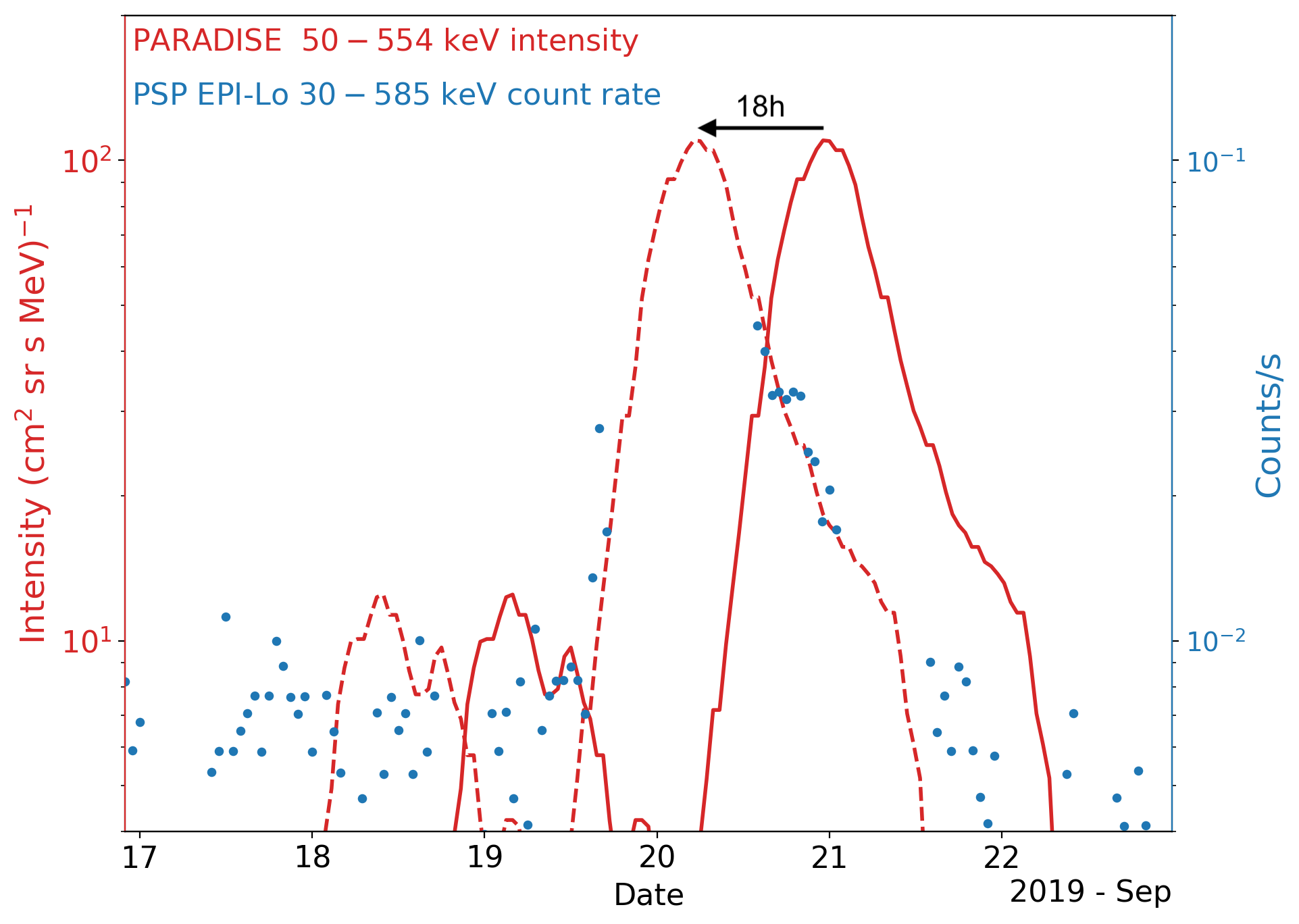}
    \caption{Count rates of 30-585 keV ions measured by the Time-of-Flight (ToF) system of the EPI-Lo instrument on PSP (blue dots) and the simulated 50-554 keV proton omnidirectional intensity (red solid line), along with the same simulated intensity but  shifted 18h in time (red dashed line) }
    \label{fig:psp_int}
\end{figure*}
Figure~\ref{fig:psp_int} shows the simulated particle intensities at PSP (red line) together with the  count rate of energetic ions measured by the 
Time-of-Flight (ToF) system of the EPI-Lo instrument \citep{mccomas16} on board PSP (blue dots). We use hourly averages accumulated over all the apertures of EPI-Lo and consider the energy ranges provided in the NASA Space Physics Data Facility\footnote{\url{cdaweb.gsfc.nasa.gov}}.
EPI-Lo data was only available for those periods plotted in Figure~\ref{fig:psp_int}, suggesting that intensity enhancements were present between 19 Sep 14:00 UT and 21 Sep 23:30~UT.

In our simulation, the omnidirectional intensities at PSP peak around 21 Sep 00:00~UT, which does not agree with the decreasing trend of the count rates observed at that time. 
However, we recall that, in the solar wind simulation, the transition from the slow to fast solar wind is inferred to be too smooth at PSP. 
In particular, the time difference between the peaks of the dashed and solid lines in Fig.\ref{fig:sw_profiles}a  is ${\sim}$18h. 
Figure~\ref{fig:psp_int} shows that by shifting the simulated particle intensity profiles by 18h (dashed red line), a qualitative agreement is obtained between the simulation and the EPI-Lo data, both at the event onset and the decay phase. 
This further confirms that the solar wind speed gradient at PSP most likely was nearly as steep as the one observed at STA.
Such a steep speed gradient indicates that particle acceleration may already have been occurring inside 0.5~au. 
However, the shifted curve in Fig.~\ref{fig:psp_int} 
shows a first peak on Sep 18 associated with the FCW, suggesting that this enhancement was likely below the background level of the instrument, and indicating that the  acceleration process at the FCW was not efficient at ${\sim} 0.5$~au.  

\subsection{Three-dimensional Proton Distributions}
\begin{figure*}
\begin{tabular}{c  c}
\includegraphics[width=0.45\textwidth]{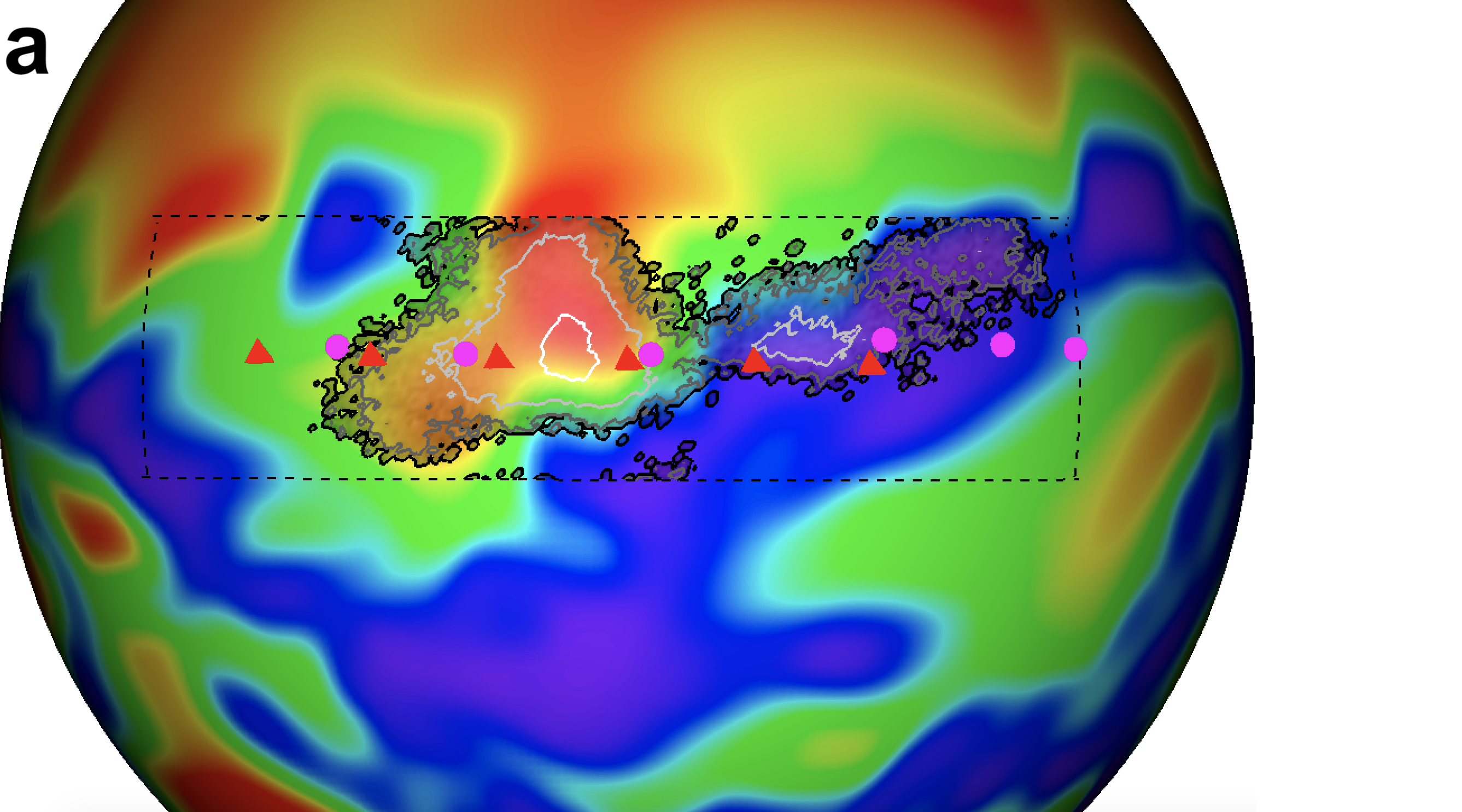} 
\includegraphics[width=0.45\textwidth]{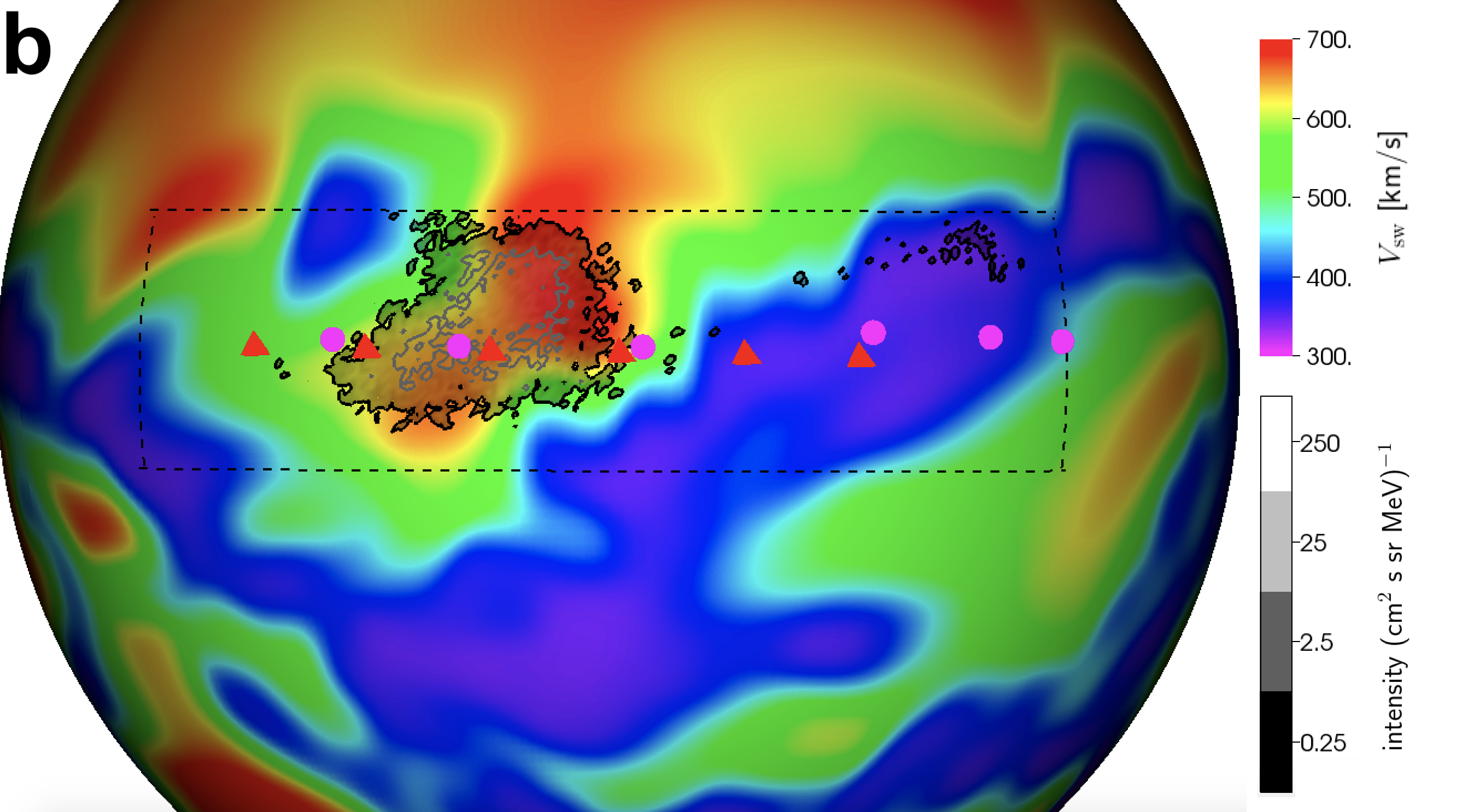}
\\

\includegraphics[width=0.45\textwidth]{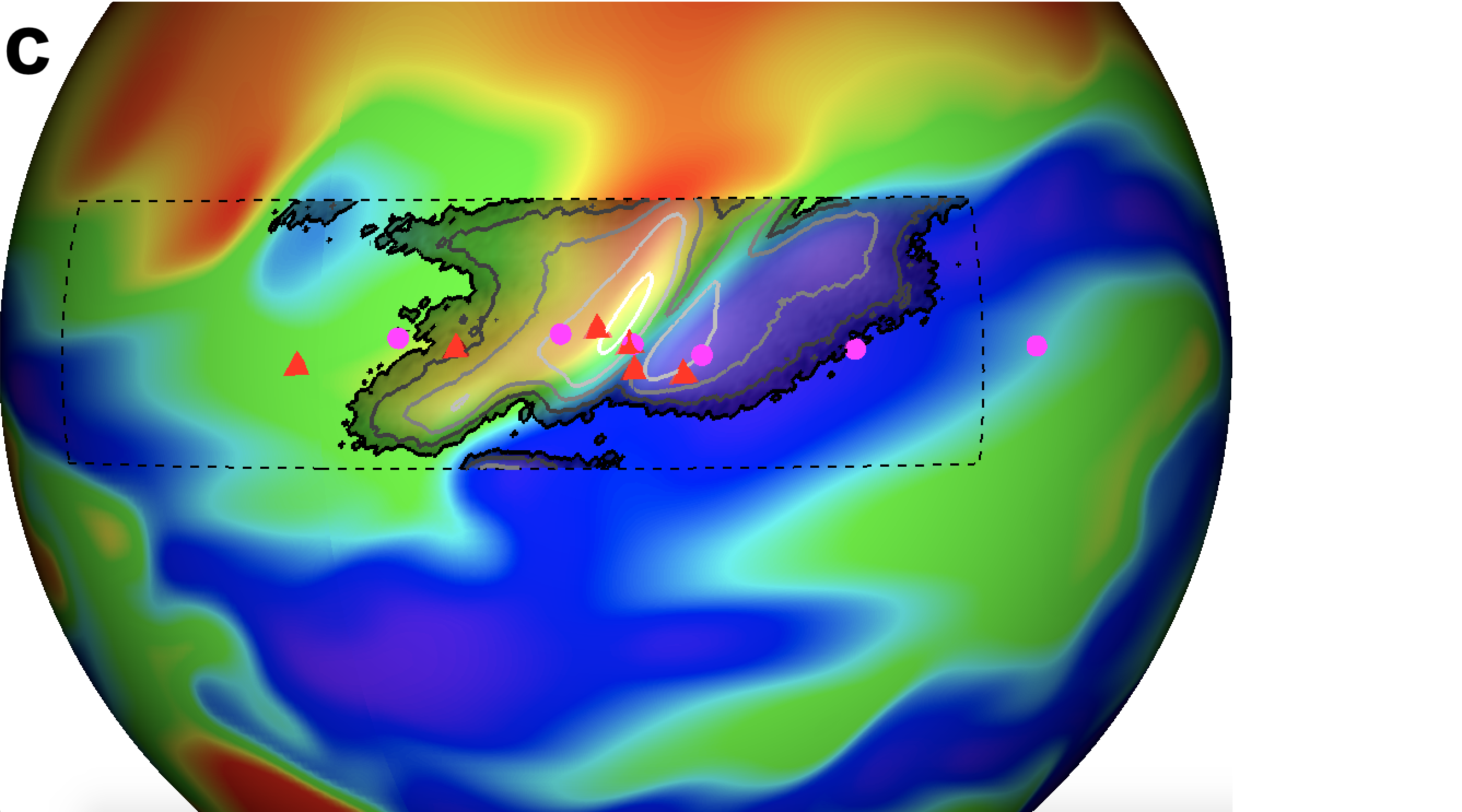} 
\includegraphics[width=0.45\textwidth]{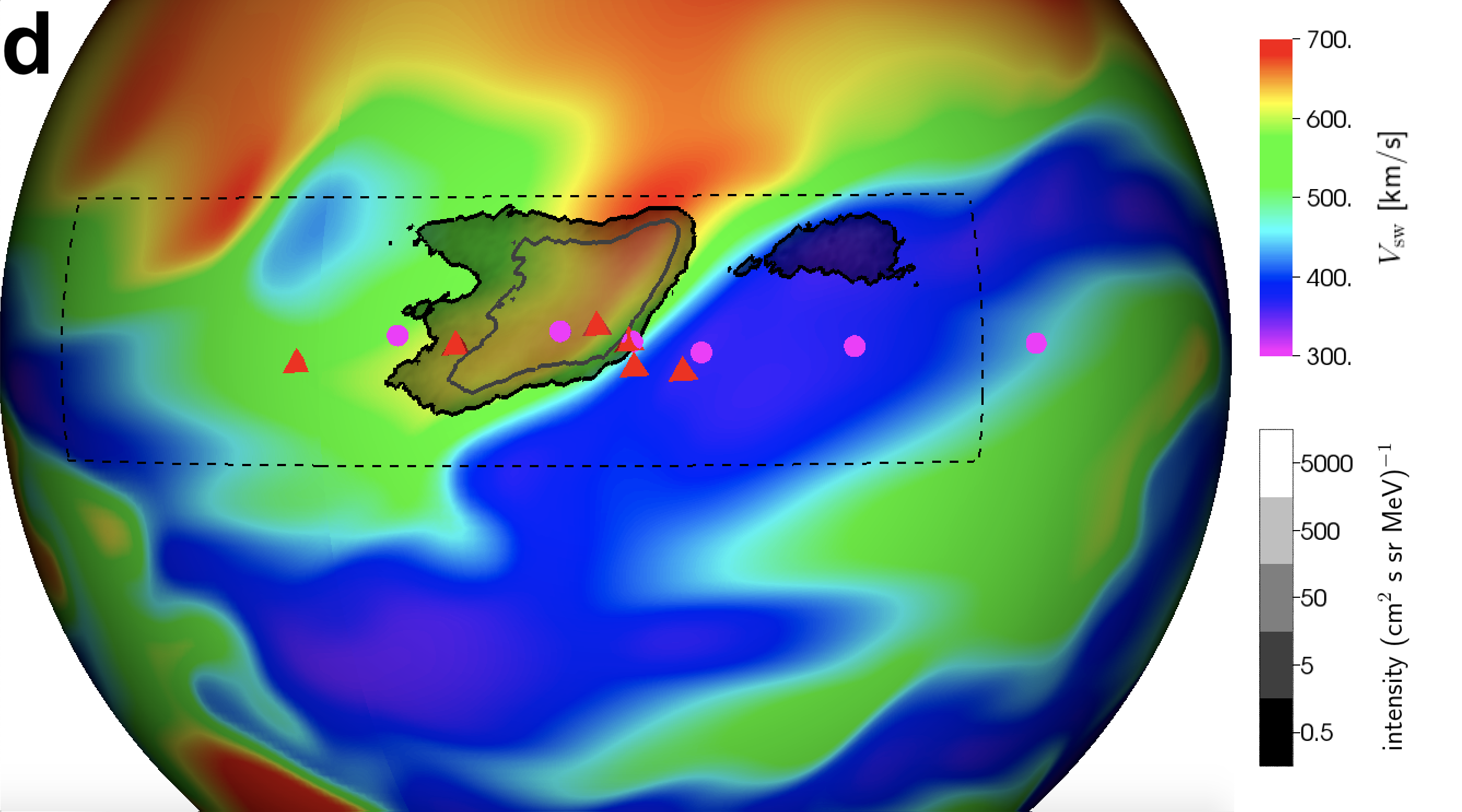}
\end{tabular}
\caption{Rainbow-colour maps showing the solar wind speed at $r = 0.5$~au (top row) and  $r = 1.5$~au (bottom row).  The grey shades show the simulated 84.1 - 92.7 keV (left) and 496.4 - 554.8 keV  (right) proton intensities. The symbols represent the magnetic connection of STEREO-A (pink dots) and PSP (red triangles) at a 24h time cadence, ranging from 18 Sep 12:00 (rightmost symbols) to 23 Sep 12:00 (leftmost symbols).  The dashed rectangles indicate the particle sampling domain of PARADISE.
}\label{fig:IntensitiesOnSphericalSlice}
\end{figure*}

Figure~\ref{fig:IntensitiesOnSphericalSlice} shows the omnidirectional particle intensity on spherical shells located at 0.5~au (top row) and 1.5~au (bottom row) for the  energy channels 84.1 - 92.7~keV (left)  and 496.4 - 554.8~keV (right).
These figures show how the spatial dependence of the particle population is strongly modulated by the shape of the HSS.
Whereas $\sim$0.5 MeV  particles are mostly found within the HSS, the $\sim$80 keV particles
extend also to the slower solar wind. At low energies, the 
maximum intensities at 1.5~au are centred around the RCW and FCW, whereas  
at 0.5~au the particles are more dispersed and two intensity peaks are attained within the slow solar wind in front of the FCW and in the fast solar wind stream behind the RCW.  
Such a spatial dependence is often observed in SIR events \citep[see][and references therein]{richardson18} and is a consequence of nonlocal particle acceleration.

Figure~\ref{fig:IntensitiesOnSphericalSlice} also illustrates how the magnetic connection of PSP (red triangles) only crosses  the southward edge of the intensity peak associated with the FCW in the simulation.  
If the spacecraft had been located at slightly higher latitude, it would have observed significantly higher particle intensities at the FCW, according to the simulation.
This might further explain why PSP did not see any particle enhancement associated with the FCW (Figure~\ref{fig:psp_int}).
The strong latitudinal dependence of the particle distribution follows from the intricate structure of the fast solar wind stream  (Figure~\ref{fig:sw_colormaps}c). 
This illustrates that two spacecraft located at slightly different latitudes can see very different particle patterns. 
This finding is in agreement with the results of \cite{jian19}, who analysed 151 pairs of SIRs seen by STEREO A and B  and showed that, even within 5$^\circ$ of latitude, the solar wind properties of a single SIR can be strongly variable.

\section{Summary and Conclusion}
In this work, we have presented the simulation results of the SIR event  observed by both PSP and STEREO-A in September 2019. 
By coupling the energetic particle model PARADISE to the MHD model EUHFORIA, we were able to model a solar wind configuration and  energetic particle enhancements that are in good agreement with the observations of STEREO-A and PSP. 
In our simulations, the energetic protons are accelerated self-consistently at the compression waves bounding the SIR, assuming a seed population of 40~keV protons. 
Such a seed population may originate from the solar wind suprathermal tail, especially near compression or shock waves where the solar wind gets heated adiabatically, thus producing more suprathermal protons \citep{prinsloo19}.
Our simulations are consistent with the hypothesis that energetic particle enhancements near SIRs are mainly produced by the diffusive acceleration of solar wind suprathermal tail particles at the compression waves bounding SIRs.
Low-energy particles ($\lesssim 500$~keV) can already accelerate at small radial distances, before the compression waves have steepened sufficiently to accelerate MeV particles. 
Therefore,  particles observed at a given heliospheric location may come from different regions, with the higher-energy particles being accelerated at larger helioradii. 
The stronger RCW (Fig.~\ref{fig:sw_colormaps}b) leads to a more efficient acceleration of high-energy particles near the trailing edge of the SIR.
For these reasons, the simulated energy spectrum hardens toward the end of the SIR event, in agreement with the observations.

Earlier models predicted that the energy spectrum associated with SIR events would exhibit a turnover at low energies ($<0.5$ MeV), because these ions undergo significant energy losses due to adiabatic deceleration as they propagate in the sunward direction \citep{fisk80}.
However, such a turnover has not been observed in recent PSP observations \citep{desai20,joyce20}.
Our simulations show that compression waves can  accelerate suprathermal particles already within the orbit of Earth, which provides an explanation on why  the energy spectra do not show a turnover at low energies. 

In addition, the strong density pileup and magnetic field enhancement observed by PSP in front of the HSS indicates that our solar wind simulation likely underestimated the speed gradient at the leading edge of the HSS within the orbit of Earth.  
Steep speed gradients at small radial distances have been previously reported in Helios data \citep{schwenn90}, suggesting a sharp transition between slow and fast solar wind sources. 
The compression waves associated with such sudden transitions might allow some particle acceleration to occur in close proximity to the Sun. 
Future PSP observations close to the Sun will allow for continued
investigation on these particle acceleration processes. 

No special turbulence conditions were assumed near the compression waves in our simulation. 
The good agreement between the energy spectrum of the simulation and the data 
suggests that any enhancements in the solar wind turbulence near the SIR might be of secondary importance to the acceleration process of  suprathermal protons.  
The particle simulations did not include a momentum diffusion process, suggesting that stochastic acceleration is likely not a dominant acceleration process inside the SIR.

Furthermore, our simulations indicate that cross-field diffusion is likely not very efficient near and within the SIR. 
This is because the start and stop times of the simulated SIR event agree well with the observed ones. 
This agreement is lost when introducing a strong cross-field diffusion process, since it produces earlier start times and later stop times. 
In addition, a strong cross-field diffusion process  would reshape the two-peak intensity profile seen in Fig.~\ref{fig:STA-intensities} into an intensity profile consisting of a single broad peak, which is not observed.

Finally, %the simulations illustrated
our simulations show that the three-dimensional configuration of the solar wind  streams  can strongly  modulate  the  energetic  particle  distributions,  illustrating the importance of employing advanced three dimensional models when studying SIR events. 
Moreover, it indicates that small latitudinal differences between spacecraft can lead to very different energetic particle observations. 
In the future, data from SolO obtained outside the solar ecliptic will provide insight into latitudinal variations of  the  particle intensities associated with SIRs, and hence allow us to further constrain the scattering properties of energetic particles.

\section*{Acknowledgements}
N. W. acknowledges funding from the Research Foundation - Flanders (FWO -- Vlaanderen, fellowship no. 1184319N).
E. S. is supported by a PhD grant awarded by the Royal Observatory of Belgium.
This project has received funding from the European Union’s Horizon 2020 research
and innovation programs under grant agreement No. 870405 (EUHFORIA 2.0).
Computational resources and services used in this work were provided by the VSC (Flemish Supercomputer Centre), funded by the FWO and the Flemish Government-Department EWI.
Support by ISSI and ISSI-BJ through the international team 469 is also acknowledged. A. A.  acknowledges the support by the Spanish Ministerio de Ciencia e Innovaci\'{o}n (MICINN) under grant PID2019-105510GB-C31 and through the ``Center of Excellence Mar\'{i}a de Maeztu 2020-2023'' award to the ICCUB (CEX2019-000918-M). D. L. acknowledges the support from the NASA-HGI grant NNX16AF73G and the NASA Program NNH17ZDA001N-LWS. The work was also carried out in the framework of the Finnish Centre of Excellence in Research of Sustainable Space (Academy of Finland Grant 312390).
These results were also obtained in the framework of the projects
C14/19/089  (C1 project Internal Funds KU Leuven), G.0D07.19N  (FWO-Vlaanderen), and C~90347 (ESA Prodex).
We acknowledge the use of STEREO SEPT data available at \url{http://www2.physik.uni- kiel.de/stereo/data/sept/}, and STEREO magnetic field and plasma data available at \url{https://stereo-ssc.nascom.nasa.gov}. We are also thankful to the Parker Solar Probe teams for providing IS$\odot$IS/EPI-Lo and magnetic field and plasma data available at \url{https://cdaweb.gsfc.nasa.gov}. Parker Solar Probe was designed, built, and is now operated by the Johns Hopkins Applied Physics Laboratory as part of NASA’s Living with a Star (LWS) program (contract NNN06AA01C).  Support from the LWS management and technical team has played a critical role in the success of the Parker Solar Probe mission.

\appendix\label{sec:methods}

\section{EUHFORIA setup}\label{app:euhforia}
The solar wind is modelled by using EUHFORIA, a physics-based coronal and heliospheric model designed for space weather research and prediction purposes \citep{pomoell18}. 
The coronal module of EUHFORIA uses as input a  synoptic  magnetogram from the Global Oscillation Network Group (GONG) and applies the semi-empirical Wang--Sheeley--Arge model \citep[WSA;][]{arge04} to provide the solar wind plasma and magnetic quantities at $0.1$~au. These data are then utilised as the inner boundary conditions for the heliospheric module. 
The latter module solves the ideal MHD equations from 0.1~au up to a prescribed outer boundary, which is placed at 4~au in this work. 
The resolution of the spatial grid on which the MHD equations are solved is chosen to be $(\Delta r, \Delta \vartheta, \Delta\varphi  )= (1.64~R_\odot, 1^\circ, 1^\circ)$, where $r$, $\vartheta$, and $\varphi$ denote the radial, latitudinal, and longitudinal coordinates, respectively. 

In this study, the GONG magnetogram of 2019-09-18 06:14UT is used as input to the coronal module and the standard setup \citep{pomoell18} of EUHFORIA is used, except for the following three changes:
\begin{enumerate}
    \item  A constant value of 30 km/s is added to the resulting WSA speed profile, and in addition, the speed profile is capped to be in the range $[340, 700]$~km/s at 0.1~au.
    \item  The WSA solar wind speed map is rotated by 7.25$^\circ$ instead of 10$^\circ$, to account for rotation of the coronal magnetic field.
    \item The magnetic field polarity is assumed to be everywhere positive  at 0.1~au.
\end{enumerate}
The first change is necessary to avoid a systematic underestimation of the slow and fast solar wind speed in the vicinity of the SIR. 
The second change is performed to ensure that the simulated arrival time of the HSS at STEREO-A corresponds with the observed arrival time.
The third change is done to avoid having a heliospheric current sheet (HCS) in the simulation, which is required by PARADISE as explained in the next paragraph.

\section{PARADISE setup}\label{app:paradise}
\begin{figure*}
    \centering
    \includegraphics[width=0.6\textwidth]{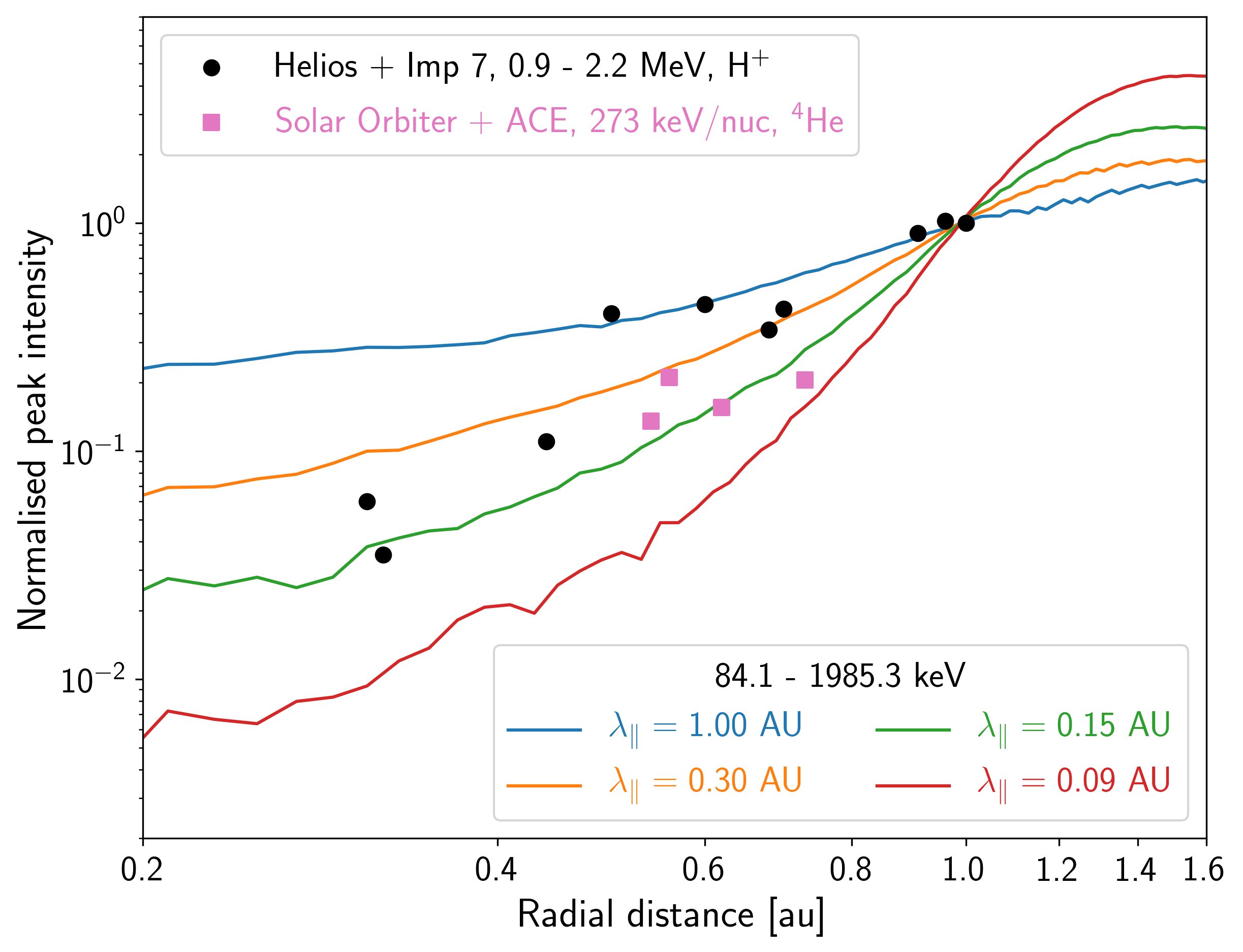}
    \caption{ Simulated 94-1985 keV proton peak intensities as a function of heliocentric radial distance for different parallel mean free paths. The intensities have been normalised to their value at 1~au.  The black dots and the grey squares correspond to SIR events observed by Helios   \citep{vanhollebeke78} and SolO \citep{allen20c}, respectively. }
    \label{fig:radial_evol}
\end{figure*}
The PARADISE \citep{wijsenPHD20} model is used to calculate the temporal and spatial evolution of  energetic particle populations that propagate through  EUHFORIA solar wind configurations.
This is achieved by solving the 5-dimensional focused transport equation  \citep[FTE; e.g.,][]{roelof69,skilling71,leRoux12}, which takes into account the effects of solar wind turbulence through a set of phase-space diffusion processes.
The FTE utilised in PARADISE is expressed in guiding centre (GC) coordinates, and contains therefore the effects of GC drifts and cross-field diffusion \citep[e.g.,][]{zhang06, wijsenPHD20,strauss20}.  
The GC drifts have, however, a negligible effect on the transport of the low-energetic protons studied in this work \citep{wijsen20}.
The FTE accounts for particles' pitch-angle and momentum changes in compressional, shear, and accelerating solar wind flows \citep[e.g.,][]{leRoux07,zank14}. 
In particular, it takes into account the acceleration of particles in high-amplitude compression and shock waves \citep{giacalone02, wijsen19a}. 

The FTE assumes that the solar wind varies on temporal and length scales that are larger than the energetic particles' gyroperiods and gryoradii, respectively. These assumptions may not be valid near the HCS \citep{burger85,wijsenPHD20}, which is why we do not include a HCS in our solar wind simulations.

In this study, 40~keV protons are injected in solar wind regions where the divergence of the solar wind velocity is negative, i.e., $\nabla\cdot\vec{V}_{\rm sw} < 0$. 
As illustrated in Fig.~\ref{fig:sw_colormaps}, these regions include the RCW and the FCW bounding the SIR. 
In addition, we scale the injected particle distribution so that $f_{\rm inj}(t,\vec{x}) \propto  |\nabla\cdot\vec{V}_{\rm sw}|$  \citep{prinsloo19}.
This is done because a negative $\nabla\cdot\vec{V}_{\rm sw}$ gives a  measure of local  compression and therefore of adiabatic heating of the solar wind. In addition, regions where $\nabla\cdot\vec{V}_{\rm sw}< 0$ are the locations where particles can gain energy by interacting with the converging scattering centres embedded in the solar wind flow.

 The pseudo-particles are propagated for a total time period of 12~days in the solar wind generated by EUHFORIA. 
A reflective boundary condition is prescribed at the inner boundary  ($0.1$~au) and an absorbing boundary condition is prescribed at the outer boundary ($4$~au).
The particle injection distribution is assumed to be constant in time in the frame corotating with the Sun.
 This is achieved by convolving a  Green's function solution of the FTE with a uniform time distribution. 
 This is possible because the EUHFORIA solar wind solution is steady in the frame corotating with the Sun.
 The resulting proton distribution obtained after 12~days is  steady-state in the corotating frame. 
 This distribution is utilised to obtain the intensity-time profiles at PSP and STEREO-A presented in this work.
 Furthermore, the distribution is scaled such that the simulated intensity matches to the observed STEREO-A (peak) intensity in the 84.1-92.7 keV channel attained  on 21 Sep 16:55~UT.

 The pseudo-particles are  sampled on a grid of phase-space volumes of size $2\pi r^2 \sin(\theta)\Delta \vartheta \Delta \varphi  \Delta r \Delta E \Delta \mu$, where $E$ denotes the energy coordinate, $\mu$ the cosine of the pitch-angle, and $(r,\vartheta,\varphi)$ are spherical coordinates. 
We choose $\Delta r = 0.02$~au, $\Delta \vartheta = \Delta \varphi = 0.5^\circ$ and $\Delta \mu = 0.1$.  
The energy $E$ is computed in the HEEQ reference frame upon sampling the pseudo-particles. 

The FTE includes the effect of turbulence on the energetic particle transport through a set of diffusion processes in phase space. 
In this work, we include a cross-field diffusion process and a pitch-angle diffusion process. 
A constant perpendicular mean free path $\lambda_\perp = 10^{-4}$~au is assumed to describe the cross-field diffusion coefficient as $\kappa_\perp = \frac{1}{3} v\lambda_\perp$.
For the pitch-angle diffusion, we prescribe an anisotropic diffusion coefficient based on Quasi-Linear theory \citep{jokipii66}, as implemented in PARADISE \citep{wijsen19a}. 
The magnitude of the pitch-angle diffusion is fixed by assuming a constant parallel mean free path $\lambda_\parallel$ everywhere in the  heliosphere.

Figure~\ref{fig:radial_evol} shows  the radial variation of the 84 - 1985 keV proton peak intensities for different parallel mean free paths. 
The intensities are calculated as a six-hour average around the peak intensity seen by virtual spacecraft that are radially aligned and located at  $2.3^\circ$ latitude (HEEQ). The intensities have been normalised to the value attained at 1~au.
The dots give the  0.96 - 2.2~MeV proton peak intensities determined by \cite{vanhollebeke78} of SIR events observed  by the Helios spacecraft, relative to the peak intensities of the corresponding SIR event observed at 1~au by Imp~7 .
The squares give the  $^4$He intensities determined by \cite{allen20c} of SIR events observed by SolO, relative to the intensities of the corresponding SIR events observed by the Advanced Composition Explorer \citep[ACE;][]{stone98} at 1~au.
Despite the differences in ion species and energy ranges, a good match is obtained between the simulations and the observations. 
Data of the Sep 2019 SIR event discussed in this work are not included in the figure, since PSP did not observe the peak intensity. 
%Due to the existence of the strong solar wind compression at PSP, we speculate that the peak intensities of this event would have been relatively high. 
A mean free path of 1.0~au produces high intensities at small radial distances, since for  such a large mean free path, the particles are only weakly coupled to the solar wind plasma.
As a result, magnetic focusing is the dominant process in reversing the propagation direction of sunward streaming particles. 
In contrast, for the mean free path of 0.09~au,  particles are more tightly coupled to the solar wind plasma, and  consequently they are efficiently advected with the solar wind toward larger radial distances,  resulting in particle intensities that decrease strongly with decreasing radial distance.  Most observations included in Fig.~\ref{fig:radial_evol} fit a parallel mean free path in the range $0.15 {\text{\, au}}\lesssim \lambda_\parallel \lesssim 0.3 {\text{\, au}}$.

Each observational point presented in Fig.~\ref{fig:radial_evol} is derived by combining the measurements of two spacecraft (Helios-IMP 7 or SolO-ACE) that were not necessarily located at exactly the same latitude.  Figure~\ref{fig:IntensitiesOnSphericalSlice} illustrates that, in our simulation, there  is a clear variation of the particle intensities with latitude 
as a result of the underlying non-nominal solar wind conditions. 
Similarly, it was shown by e.g., \cite{schwenn78} and \cite{jian19} that spacecraft separated in latitude by just a few degrees can observe significant differences in the plasma properties of a single SIR.
Such a latitudinal variation might be prevalent in SIR events and will contribute to the dispersion of the observational points presented in Fig.~\ref{fig:radial_evol}.  
In order to infer any general latitudinal intensity dependencies of SIR events, detailed studies of various SIR events seen at different latitudes are needed.

For the Sep 2019 SIR event discussed in this work, the best agreement between the simulated and observed event-integrated energy spectrum at STEREO-A is attained for $\lambda_\parallel = 0.3$~au. The latter mean free path also provides the best agreement between the observed and simulated start and stop times of the SIR event at both  STEREO-A and PSP. Therefore, the simulation results presented in this work are for $\lambda_\parallel = 0.3$~au.

%\bibliography{allpapers}

\end{document}